\documentclass[12pt]{iopart}

\usepackage{graphicx}
\usepackage[usenames]{color}
\newcommand{\beq}{\begin{equation}}
\newcommand{\eeq}{\end{equation}}
\newcommand{\beqa}{\begin{eqnarray}}
\newcommand{\eeqa}{\end{eqnarray}}
\newcommand{\ba}{\begin{array}}
\newcommand{\ea}{\end{array}}

\begin{document} \title[Dynamics of bright and vortex solitons of a 
dipolar BEC]{Dynamics of quasi-one-dimensional bright and vortex 
solitons of a dipolar Bose-Einstein condensate with repulsive atomic 
interaction}

\author{Luis E. Young-S.$^1$,
P. Muruganandam$^{1,2}$\footnote{anand@cnld.bdu.ac.in},
and S. K. Adhikari$^1$\footnote{Email: adhikari@ift.unesp.br; URL:
http://www.ift.unesp.br/users/adhikari/}}
\address{$^1$Instituto de F\'{\i}sica Te\'orica, UNESP - Universidade Estadual Paulista, 01.140-070 S\~ao Paulo, S\~ao Paulo, Brazil}
\address{$^2$School of Physics, Bharathidasan University, 
Palkalaiperur Campus, 
Tiruchirappalli 620024,   
Tamilnadu, 
India}

\begin{abstract} By numerical and variational analysis of the 
three-dimensional Gross-Pitaevskii equation we study the formation and 
dynamics of bright and vortex-bright solitons
 in a cigar-shaped dipolar 
Bose-Einstein condensate  for 
large {\it repulsive}
atomic interactions. 
Phase diagram showing the region of stability of the  
solitons is obtained. 
We also study the dynamics of breathing 
oscillation of the solitons 
as well as the collision dynamics of  two 
solitons at large velocities. 
Two solitons placed side-by-side at rest coalesce to form a stable bound 
soliton molecule
due to dipolar attraction.
 \end{abstract}


\maketitle

A bright soliton is a self-reinforcing solitary wave  
 that maintains its shape, while traveling at constant speed, due 
to a cancellation of nonlinear attraction and dispersive effects. 
Experimentally, bright matter-wave solitons and soliton trains were 
created in a Bose-Einstein condensate (BEC) of $^7$Li \cite{4r,4rb} 
and $^{85}$Rb atoms \cite{5r} by turning the atomic interaction attractive from 
repulsive using  a Feshbach 
resonance (FR) \cite{fesh} and releasing the BEC in an axially free 
or an expulsive 
trap. 

Lately, 
a BEC of $^{52}$Cr atoms 
with a large long-range
dipolar interaction has been made \cite{pfau,rpp}. This allows  to study the
dipolar BEC (DBEC)
 with variable  short-range interaction \cite{pfau,rpp}
using a FR \cite{fesh}.
The DBEC has many
distinct features \cite{pfau,jb,Ronen2007,Dutta2007}. 
The stability of a DBEC depends not only 
on the scattering length, but also on the trap geometry 
 \cite{pfau,jb,Dutta2007}. A
disk-shaped trap 
gives a repulsive dipolar interaction 
 and  the DBEC is
more stable. In contrast, a cigar-shaped trap 
yields an attractive dipolar interaction
 and hence may favor a collapse
\cite{pfau,Dutta2007,Santos2000}. {
Collapse instability of solitons in the nonpolynomial Schr\"odinger equation with dipole-dipole interactions was 
studied by Gligori\'c {\it et al.} \cite{giloric}.}
The controllable
short-range interaction  with the exotic 
dipolar interaction 
makes the DBEC an interesting system for
soliton generation in the cigar configuration due to added 
{\it dipolar attraction}
and a challenging system for
theoretical investigation \cite{Giovanazzi2002}
to study the interplay between the
dipolar  and  short-range interactions.

We suggest 
cigar-shaped  bright and vortex 
solitons \cite{vortex} for {\it repulsive} short-range interaction
in a DBEC free to move along the axial direction and 
 trapped in the radial direction, using the three-dimensional (3D) 
Gross-Pitaevskii (GP) equation, and study their statics and dynamics. 
Vortex-bright solitons (called vortex solitons hereafter)
are solitons of (quantized) unit angular momentum.
{In the usual
(non-dipolar) BEC, solitons with intrinsic vorticity trapped
in the cigar-shaped potential, were considered by Salasnich
{\it et al.} \cite{salasnich}.} 
Vortex states of higher angular momentum ($l>1$) are unstable 
and decay into multiple vortices of angular momentum $\hbar$ \cite{multiple}.  
For a conventional BEC 
without dipole moment, solitons appear only for attractive atomic 
interaction. Phase plots showing the stability of the solitons 
for different dipolar and short-range interactions are obtained. We 
study  breathing oscillation and collision of two bright and 
two 
vortex solitons as well as  the statics and dynamics of the bright 
solitons 
using a  Gaussian variational approach. 
{
Collision of two bright and vortex solitons in non-dipolar BEC was studied in \cite{adhinjp}. }
Where applicable, 
the variational results are  found to be 
in good agreement with numerical 
results. 
By reversing the sign of the dipolar interaction, Pedri and 
Santos \cite{Pedri2005} studied stable two-dimensional (2D) 
bright solitons in a disk-shaped DBEC free to move in the radial plane 
and 
trapped in the axial direction. {These solitons were 
later 
generalized to 
 2D 
 vortex solitons \cite{Tikhonenkov2008pra}. There has also been study of 
anisotropic 2D solitons 
not requiring the 
sign-reversal of the dipolar interaction 
\cite{Tikhonenkov2008}. } 
However, special 
conditions are needed to prepare these
 2D solitons, whereas 
the present one-dimensional (1D) solitons  can be formed under very general conditions 
and hence, are of 
greater experimental interest. { Solitons in the Tonks-Girardeau gas with dipolar interaction
was studied by Baizakov {\it et al.} \cite{baizakov} in a 1D model.}

\begin{figure}
\begin{center}
\includegraphics[width=.7\linewidth]{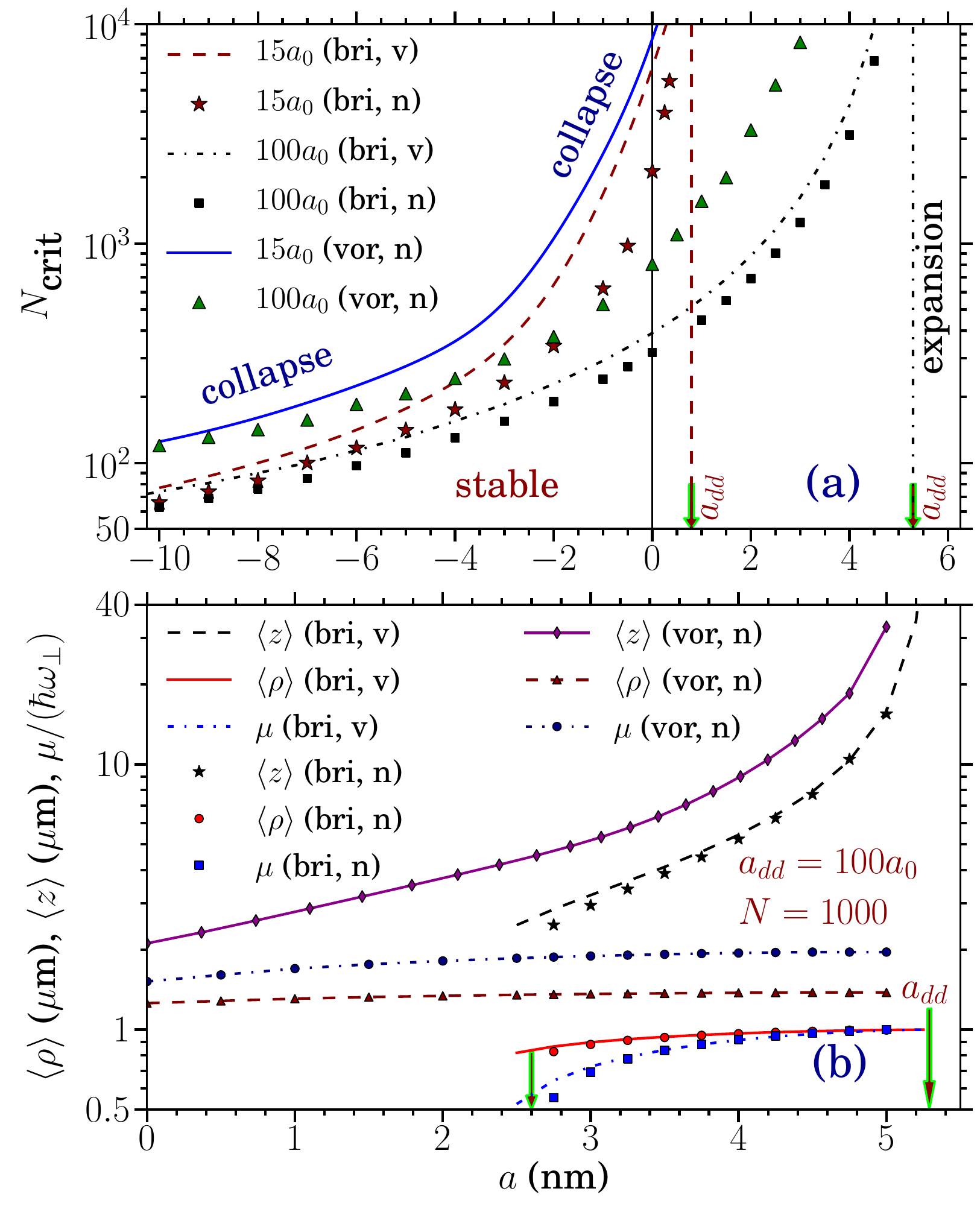}
\end{center}

\caption{(a) $N_{\mathrm{crit}}$ versus  $a$ plot 
for  
$a_{dd}=15a_0,$ $100a_0$.  (b)  rms sizes $\langle \rho \rangle, \langle z \rangle$ and chemical 
potential $\mu$ versus $a$ for  $N=1000$ and $a_{dd}
=100a_0$. The two arrows denote the region of stability of bright solitons.
(variational: v, numerical: n, bright: bri, vortex: vor.)
}
\label{fig1}
\end{figure}

{
Cuevas {\it et al.}  \cite{cuevas} studied solitons in cigar-shaped DBEC 
using a reduced quasi-1D toy model \cite{red1d}. 
As there is no collapse in 1D (models with cubic nonlinearity as in this case), 
such study predicts soliton 
for attractive interaction for an arbitrarily large number of atoms. But non-dipolar  
BEC bright solitons  \cite{perez} and vortex solitons \cite{pre}
in 3D are stable upto a critical number of atoms, 
beyond which the system collapses. 
The same is found to be true for a 
DBEC; viz. the critical numbers reported in 
figure  \ref{fig1} (a) DBEC soliton.   }


We study the bright  solitons  
in a  
DBEC of $N$ atoms, each of mass $m$, using the  GP
equation: \cite{pfau}
\begin{eqnarray}  \label{gp3d} 
i \frac{\partial \phi({\bf r},t)}{\partial t}
& =&  \biggr[ -\frac{1}{2}\nabla^2 +\frac{\rho^2}{2} + 4\pi a N|\phi({\bf r},t)|^2\nonumber \\
&+& N\int U_{dd}({\bf r -r'})|\phi({\bf r'},t)|^2d{\bf r'}
\biggr] \phi({\bf r},t), \end{eqnarray} 
with the dipolar interaction 
$ U_{dd}({\bf R}) = 3
a_{dd}(1-3\cos^2\theta)/R^3,\quad {\bf R=r-r'}$
and  normalization $\int \phi({\bf r})^2 d {\bf r}$ = 1.
 Here,   $a$ is the atomic scattering length, $\theta$ 
the angle between $\bf R$ and the polarization direction $z$.  
The length $a_{dd}
=\mu_0\bar \mu^2 m /(12\pi \hbar^2)$ 
measures 
the strength of 
dipolar interaction and its experimental
value for $^{52}$Cr  is $15a_0$ \cite{pfau}, with  $a_0$ the Bohr 
radius, 
 $\bar \mu$ the (magnetic) dipole moment of a single atom, and $\mu_0$ 
the permeability of free space. The trap $\rho^2/2$ 
acts in the radial $\rho \equiv (x,y)$ direction only. 
To obtain a quantized vortex of angular momentum $l\hbar, l=1,$ 
around axial $z$ axis, one has to introduce a phase (equal to the 
azimuthal angle)
in the wave function \cite{dst}. 
This procedure 
introduces a centrifugal term $l^2/2\rho^2$ in the GP equation for 
a vortex  and we adopt this method to study  
vortex solitons.
In  (\ref{gp3d}) we use oscillator unit of transverse trap: 
unit of time is inverse
angular frequency  $\omega_\perp$ taken as 1 ms,
unit of length is oscillator length $a_\perp\equiv 
\sqrt{\hbar/m\omega_\perp}$ taken as 1 $\mu$m.



Lagrangian density of 
  (\ref{gp3d}) for bright soliton   is \cite{vortex}
\begin{eqnarray}
{\cal L}=& \,\frac{i}{2}\left( \phi \phi^{\star}_t
- \phi^{\star}\phi_t \right) +\frac{1}{2}\vert\nabla\phi\vert^2
+ \frac{\rho^2}{2}|\phi|^2+ 2\pi aN\vert\phi\vert^4
\nonumber \\ & \,
+ \frac{N}{2}\vert
\phi\vert^2\int U_{dd}({\mathbf r}-
{\mathbf r'})\vert\phi({\mathbf r'})\vert^2 d{\mathbf r}'
.\label{eqn:vari}
\end{eqnarray}
For a variational study we
use the Gaussian ansatz  \cite{you}: 
$ \phi({\bf
r},t)=
\exp(-
{\rho^2}/{2w_\rho^2}- {z^2}/{2w_z^2}$ $ +i\alpha\rho^2
+i\beta z^2 )  /({w_\rho \sqrt w_z}\pi^{3/4})$
where $w_\rho$ and $w_z$ are time-dependent widths and 
$\alpha$ and $\beta$ are time-dependent phases. 
The
effective Lagrangian $L$ (per particle) is
\begin{eqnarray}
L &\equiv  &  \int {\cal L}\,d{\mathbf r}
 =  \left(w_\rho^2\dot{\alpha} +\frac{1}{2}
w_z^2\dot{\beta}+2w_\rho^2\alpha^2+w_z^2\beta^2          \right) 
\nonumber \\ 
&+&
\frac{1}{2}\bigg(\frac{1}{w_\rho^2} + \frac{1}{2w_z^2}
+ w_\rho^2 \bigg) + {\cal E}_{\mathrm{dip}}, 
\end{eqnarray} 
with ${\cal E}_{\mathrm{dip}}= N[a-a_{dd}f(\kappa)]/(\sqrt{2 \pi}w_\rho^2w_z)  , f(\kappa)=[1+
2\kappa^2-3\kappa^2d(\kappa)]/
(1-\kappa^2), d(\kappa)
=(\mbox{atanh}\sqrt{1-\kappa^2})/\sqrt{1-\kappa^2}, \kappa=w_\rho/w_z.$
In the cigar shape, the dipolar interaction becomes attractive and contributes
 negatively in $\cal E_{\mathrm{dip}}$.    
The Euler-Lagrange equations for parameters 
$ w_\rho, w_z, \alpha, \beta$
can be used to obtain the following equations 
for 
the widths 
\begin{eqnarray} \label{f3}  &&
\ddot{w}_{\rho}+  {w_\rho}=
\frac{{{1}}}{w_\rho^3} +\frac{
1}{\sqrt{2\pi}} \frac{N}{w_\rho^3w_{z}}
\left[2{a} - a_{dd}{g(\kappa) }\right],
\label{eq:dimless:a} \\ && \ddot{w}_{z}  =
\frac{1}{w_z^3}+ \frac{ 1}{\sqrt{2\pi}}
\frac{2N}{w_\rho^2 w_z^2} \left[{a}-
a_{dd}h(\kappa)\right], \label{f4} \end{eqnarray}
with $g(\kappa)=[2-7\kappa^2-4\kappa^4+9\kappa^4
d(\kappa)]/(1-\kappa^2)^2, h(\kappa)= [1+10\kappa^2-2\kappa^4-9\kappa^2
d(\kappa)](1-\kappa^2)^2.$
Equations (\ref{f3}) and (\ref{f4}) determine the dynamics of bright solitons. 
The stationary widths are obtained from these equations by setting 
$\ddot w_\rho=\ddot w_z=0$. 
The stationary  chemical potential ($\mu$) is given by 
$\mu=1/2w_\rho^2+1/4w_z^2+w_\rho^2/2+2{\cal E}_{\mathrm{dip}}.$

We perform a 3D numerical simulation in Cartesian $x,y,z$ variables 
employing imaginary- and real-time propagation with Crank-Nicolson 
method using Fortran programs provided in Ref. \cite{Muruganandam2009}.  
The dipolar interaction is
evaluated by the usual fast Fourier transformation  \cite{jb}.

For bright solitons the system must be attractive. In the absence of 
dipole moment ($a_{dd}=0$), attraction corresponds to $a<0$.  For 
$a_{dd}>0$, the dipolar interaction in  (\ref{gp3d}) contributes 
attractively in the cigar shape, which can compensate for 
the atomic short-range repulsion, thus forming bright solitons for 
$a>0$. From  (\ref{f4}), the onset of attraction for  a 
bright soliton is at $[a-a_{dd}h(\kappa)]=0$ while one must 
have $N\to \infty$. Then from  (\ref{f3}), as $N\to \infty$, one must 
have $[2a-a_{dd}g(\kappa)]=0$. These two conditions can be 
simultaneously satisfied as $\kappa \to 0$ for a weakly bound soliton of 
infinite extension along axial direction while 
$h(\kappa=0)=g(\kappa=0)/2=1$. Consequently, bright solitons are allowed 
 for $-\infty<a<a_{dd}$ and for $N$ below a critical number 
$N_{\mathrm{crit}}$, as  (\ref{f3}) and (\ref{f4}) only permit 
solution for $N<N_{\mathrm{crit}}$. The system collapses for 
$N>N_{\mathrm{crit}}$.  These results are illustrated in figure 
\ref{fig1} (a), where we show $N_{\mathrm{crit}}$ versus $a$ for 
$a_{dd}= 15a_0$ ($^{52}$Cr atom), and $100a_0$, for 
bright and vortex solitons from numerical and variational calculation. 
The value $a_{dd}=100a_0$ is considered to simulate a possible DBEC of 
dipolar atoms 
(or molecules) of larger dipole moment, such as $^{164}$Dy$_{66}$, 
or $^{166}$Er$_{68}$,  which are being considered for BEC experiment
 \cite{dy,dy2}. 
Preliminary study \cite{dy2}
indicates that the dysprosium atom has 
a dipole moment of
about nine times larger than that of $^{52}$Cr, which 
justifies the use of  $a_{dd}=100a_0$.   
The 
attractive region of collapse ($N>N_{\mathrm{crit}}, a<a_{dd}$), stable 
soliton formation ($N<N_{\mathrm{crit}}, a<a_{dd}$) and the repulsive 
region of expansion ($a>a_{dd}$) are clearly shown in figure 
\ref{fig1} (a).  Due to 
the centrifugal repulsion, the vortex soliton can accommodate a larger 
number of atoms \cite{pre}.  In our numerical simulation we shall 
consider DBEC 
with  $a_{dd}>a>0$ $-$ domain of soliton controlled 
solely by dipolar interaction and inaccessible for a normal BEC with 
$a_{dd}=0$. In figure \ref{fig1} (b) we plot 
 results for root-mean-square (rms) sizes 
$\langle \rho \rangle, \langle z \rangle$ and chemical potential $\mu$ 
of bright and vortex solitons.
The variational results for bright solitons  
are in good agreement with the numerical results. 
From figures \ref{fig1} (a) and (b) we find that the vortex soliton is stable in 
a wider domain of parameter space, due to the vortex core which keeps the 
atoms apart to avoid the collapse. Consequently, the vortex soliton is larger  
in size than the bright soliton with a larger chemical potential.

\begin{figure}
\begin{center}
\includegraphics[width=\linewidth]{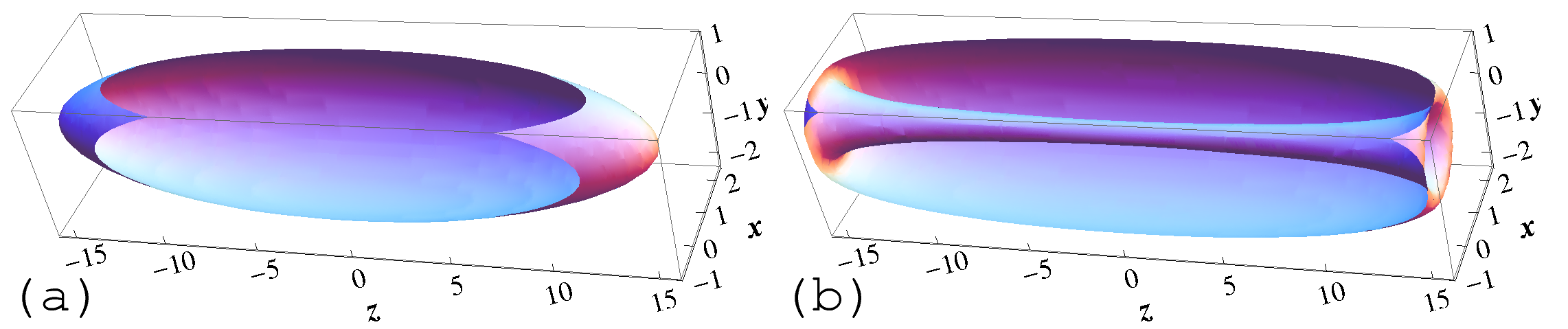}
\end{center}

\caption{ A sectional view of the 
 3D contour plot of density for a (a)
bright ($a_{dd}=15a_0$ and $a=0.5$ nm)
 and (b) vortex ($a_{dd}=100a_0$,  $a=4$ nm, $l=1$)
solitons  of  $N=1000$ atoms. 
The density on the contour is 0.001. 
}
\label{fig2}
\end{figure}

To illustrate the shape of the quasi-1D soliton, in figure \ref{fig2} 
(a) we show a section of the 3D contour plot of density of a bright 
soliton of 1000 atoms for $a=0.5$ nm, and $a_{dd}=15a_0= 0.7938$ nm.  A 
similar plot for the vortex soliton for $N=1000, a=4$ nm, and $a_{dd}=100a_0= 
5.2917$ nm is shown in figure \ref{fig2} (b) which clearly exhibits the 
vortex core along the $z$ axis.

\begin{figure}[!b]
\begin{center}
\includegraphics[width=.7\columnwidth]{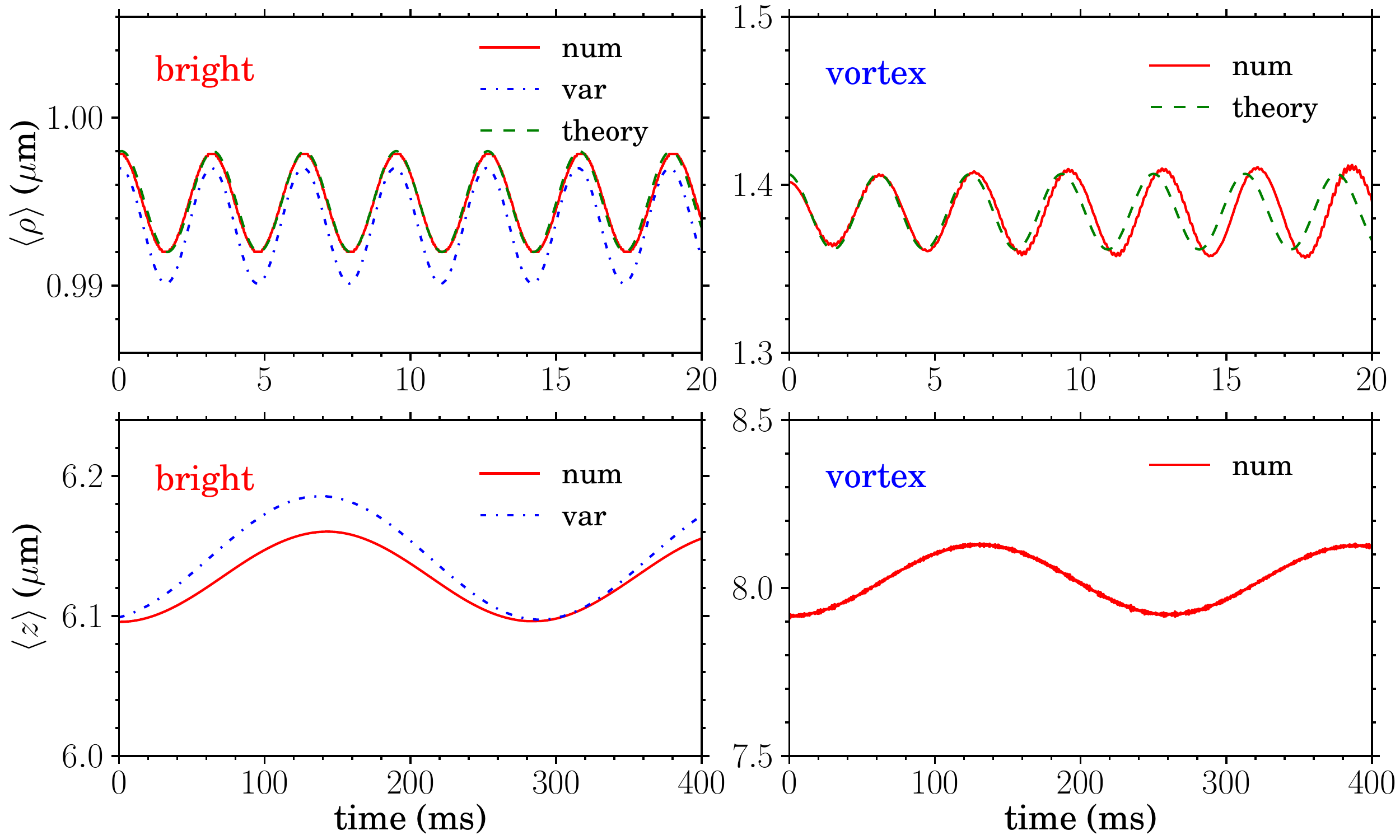}
\end{center}

\caption{ rms sizes $\langle z\rangle$ (lower panels)
and $\langle \rho 
\rangle$ (upper panels)
versus time during breathing oscillation of the bright (left panels) and 
 vortex (right panels) solitons of figure \ref{fig2}
started by jumping  $a$ from 0.5 to 0.5025 nm and from 4 to 4.02 nm, 
respectively: numerical (num), 
variational \cite{stringari} (var),  theoretical (theory).
}
\label{fig3}
\end{figure}

Next we study by numerical (num)
and variational (var)
approaches
the stability of the bright and vortex solitons shown in figure
\ref{fig2}
under 
small breathing oscillation started by a sudden change in  
scattering length $a$. This can be implemented experimentally 
by the FR technique.
 The initial soliton  was 
created by imaginary-time propagation and subsequent 
dynamics generated by real-time propagation. In figure \ref{fig3} we plot 
 rms sizes $\langle z\rangle$ and $\langle \rho \rangle$ versus time. 
  For the breathing oscillation of $\langle \rho \rangle$, we also plot 
a model result calculated from the theoretical frequency of twice the 
 trap frequency $\omega _\perp$ for an extreme 
cigar-shaped condensate using a 
hydrodynamic description \cite{stringari}. For the bright soliton, the 
angular frequencies for $\langle \rho \rangle$ are 
$\omega_\rho=2\omega_\perp$ (num,var, and theory). For the vortex 
soliton, $ \omega_\rho=1.94\omega_\perp$ (num) and $2\omega_\perp$ 
(theory). The angular frequency for $\langle z \rangle$,  controlled by the 
nonlinearity and not by any external trap, is very small in both cases.

\begin{figure}
\begin{center}
\includegraphics[width=.7\columnwidth]{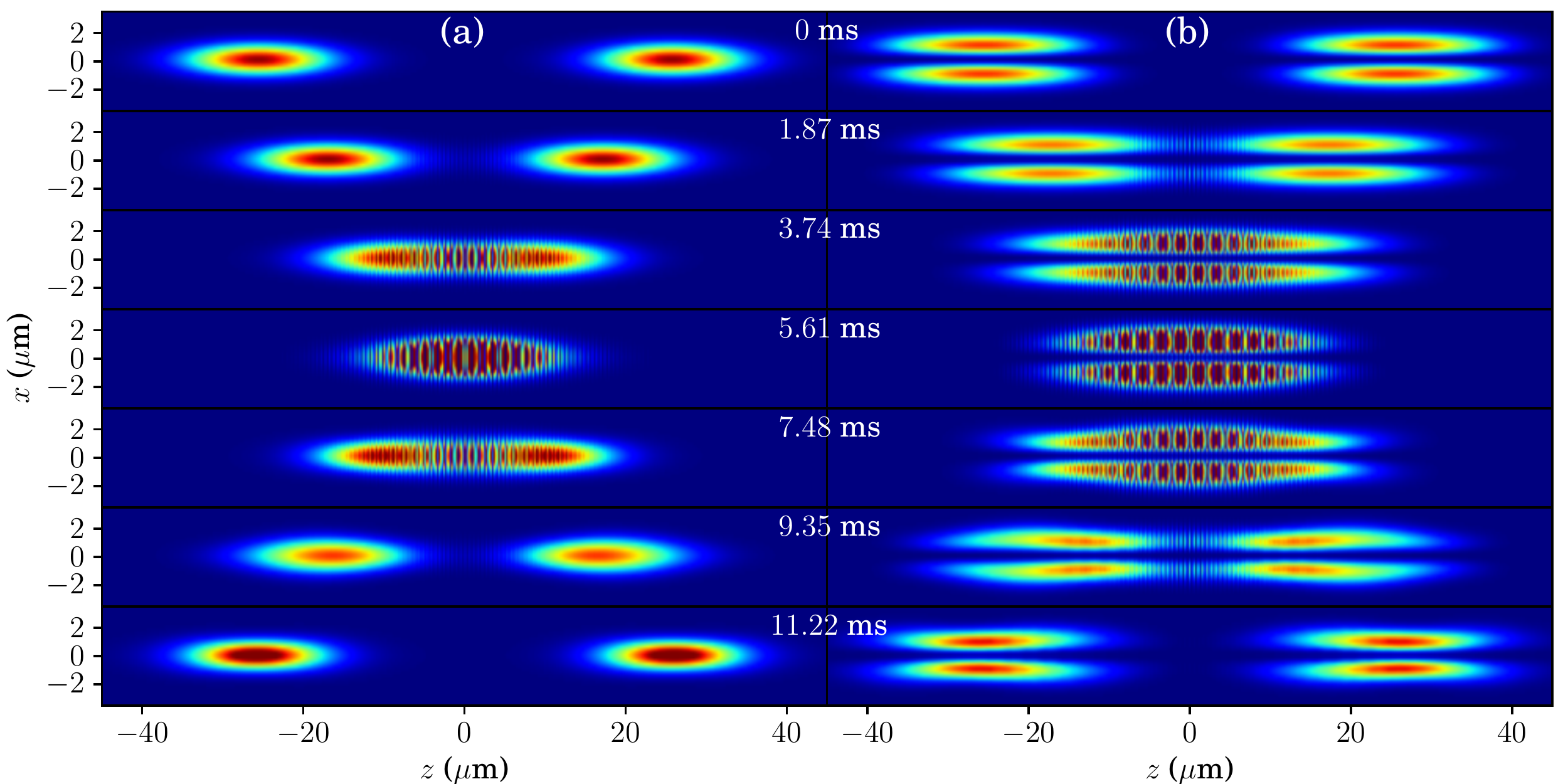}
\end{center}

\caption{  
Contour  plot of density $|\phi(x,0,z,t)|^2$
of two colliding (a) bright  
and (b) vortex 
solitons of figure \ref{fig2} with   relative velocity 1 cm/s,    
before, during and after collision. 
}
\label{fig4}
\end{figure}

Now  we investigate  the collision between two bright and two vortex 
solitons  of figure \ref{fig2}.
Two such solitons  
are placed at $z=\pm 25.6$ at $t=0$ and are then advanced 
by real-time propagation of   (\ref{gp3d}) with 
$N=2000$. Each soliton is attributed a velocity of about 5 mm/s towards 
center $z=0$ by including a phase factor $\exp(ivz)$ with $v$ a 
constant  in the initial wave function. The real-time simulation
is terminated when the solitons reach $z=\pm 25.6$ at $t=11.22$ ms. 
The 
collision dynamics is illustrated in figure \ref{fig4} for bright and vortex 
solitons, where we show snapshots of contour plots of density 
 $|\phi(x,0,z,t)|^2$ at different times before (0, 1.87 ms, 3.74 ms), 
during (5.61 ms) and after (7.48 ms, 9.35 ms, 11.22 ms) collision. The 
solitons come towards each other, interact at $z=0$ and then separate 
and come out practically unchanged.  The 
quasi-elastic collision at such a large relative velocity of 1 cm/s 
demonstrates the solitonic nature and robustness of the bright and 
vortex solitons. 
{ Video clips  of the collision dynamics of figure \ref{fig4}
are also prepared for bright 
and vortex
solitons  and contained in supplementary clips S1 {\color{red}(also at http://www.youtube.com/watch?v=ds2CzZqN9M4)} 
and S2 {\color{red}(also at http://www.youtube.com/watch?v=aQgLLhW08mE)}, respectively.  }

\begin{figure}
\begin{center}
\includegraphics[width=\linewidth]{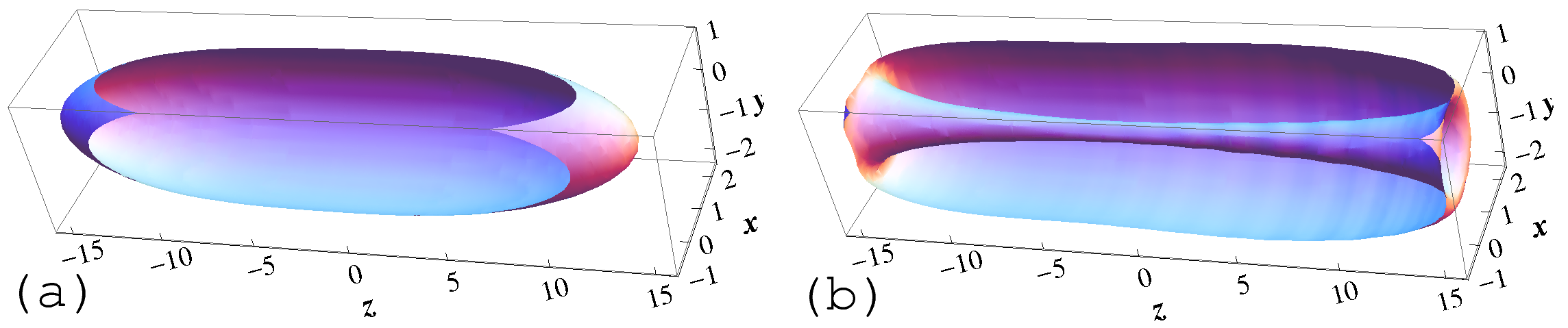}
\end{center}

\caption{  Same as figures \ref{fig2} (a) and (b) 
at time $t=11.22   $ ms after the collision 
considered in figure 
\ref{fig4}. 
}
\label{fig5}
\end{figure}

To illustrate  the quasi-elastic nature of collision, we show  in 
figures \ref{fig5} (a) and (b) the 3D contour density plot  of the final 
states  at time $t= 11.22$ ms 
after the collision  illustrated in figure 
\ref{fig4}  for bright and vortex solitons, respectively.  
The similarity of plots  in figures \ref{fig5} (a) and 
(b) for the final states after collision with the initial  states  
  in figures \ref{fig2} (a) and (b) demonstrates the elastic nature  of the 
collision. Specially, in figures  \ref{fig2} (b)  and  \ref{fig5}  (b) the radius 
of the vortex cores are practically the same determined by the fundamental 
healing length of the DBEC.    
To quantitatively compare the solitons 
before ($t=0$) and after ($t=11.22$ ms)
collision, we show in Table I  the numerical and variational 
results of 
chemical potential,
rms  sizes, and number of atoms 
 of the initial and final solitons, in good 
agreement with each other. There is no exchange of atoms between solitons
during collision.

\begin{table}
\label{I}
\caption{Numerical (n) and variational (v)  $\mu$, $\langle z\rangle$,
$\langle \rho \rangle$, and $N$   
 of the solitons of figures \ref{fig2} and  \ref{fig5}
at $t=0$ and 11.22 ms.}

\centering
\label{table:1}
\begin{tabular}{cccccc}
\hline
 & bright (v)&bright  (n)  & bright (n)  & vortex (n) & vortex  (n) \\
$t$ & 0    &0  & 11.22 ms  & 0  &  11.22 ms  \\
\hline
$\mu$ &  0.9815&   0.9822 & 0.9877 &  1.9686 & 1.9779  \\
$\langle z\rangle$ &6.0973 & 6.0958 & 6.0805 & 7.9749 &8.0088  \\
$\langle \rho \rangle$ &0.9935  &0.9924 & 0.9625 &  1.4018 &1.4030  \\
$N$ & 1000&1000 & 1000 & 1000 &1000  \\ \hline
\end{tabular}
\end{table}

\begin{figure}
\begin{center}
\includegraphics[width=.7\columnwidth]{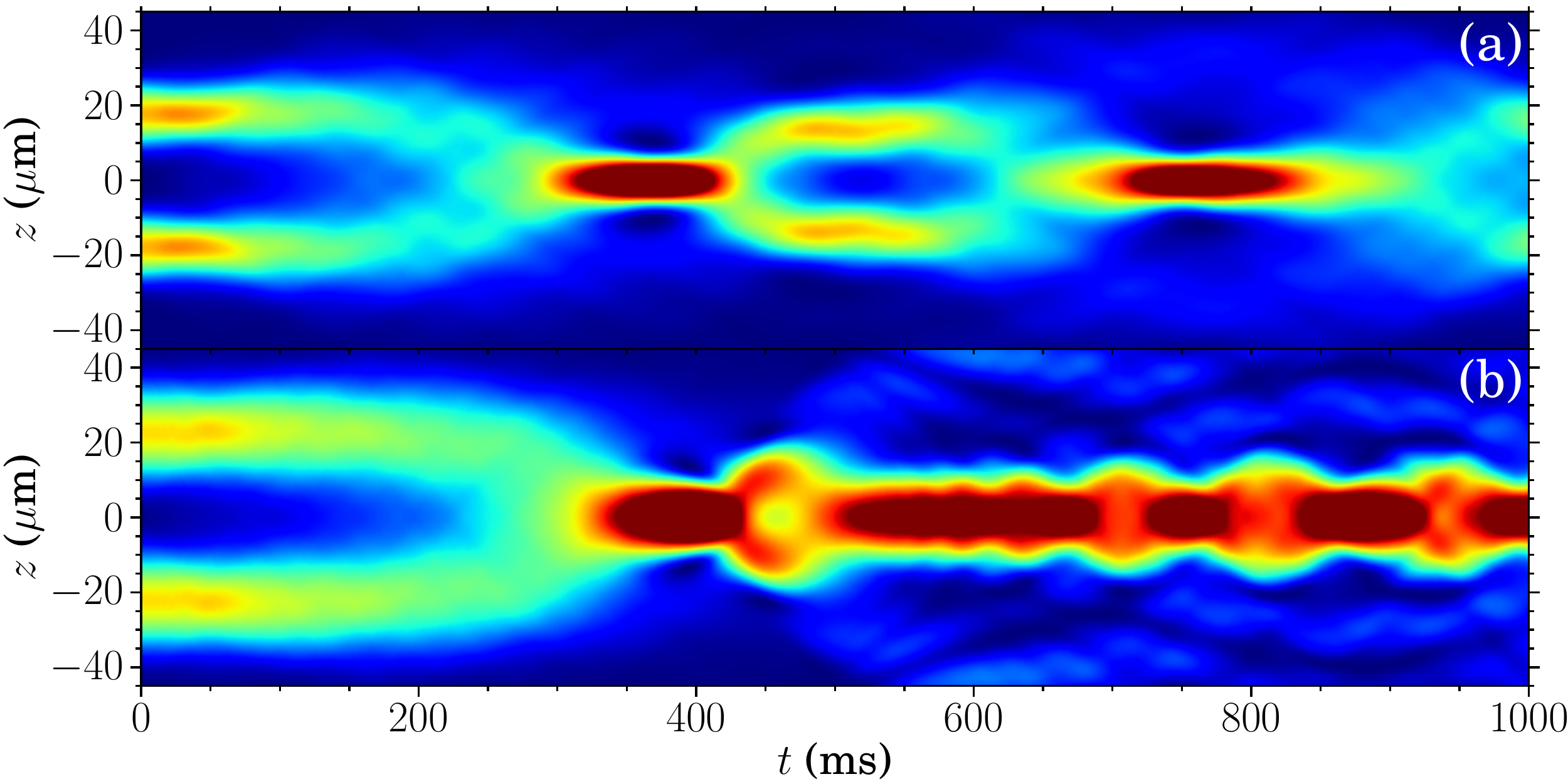}
\end{center}

\caption{ Dynamics of soliton pair formation  
from contour plot of 
1D density 
versus time 
of (a) two  bright 
and (b) vortex  
solitons of figure \ref{fig2} placed side by side at rest. 
}
\label{fig6}
\end{figure}

\begin{figure}
\begin{center}
\includegraphics[width=.7\columnwidth]{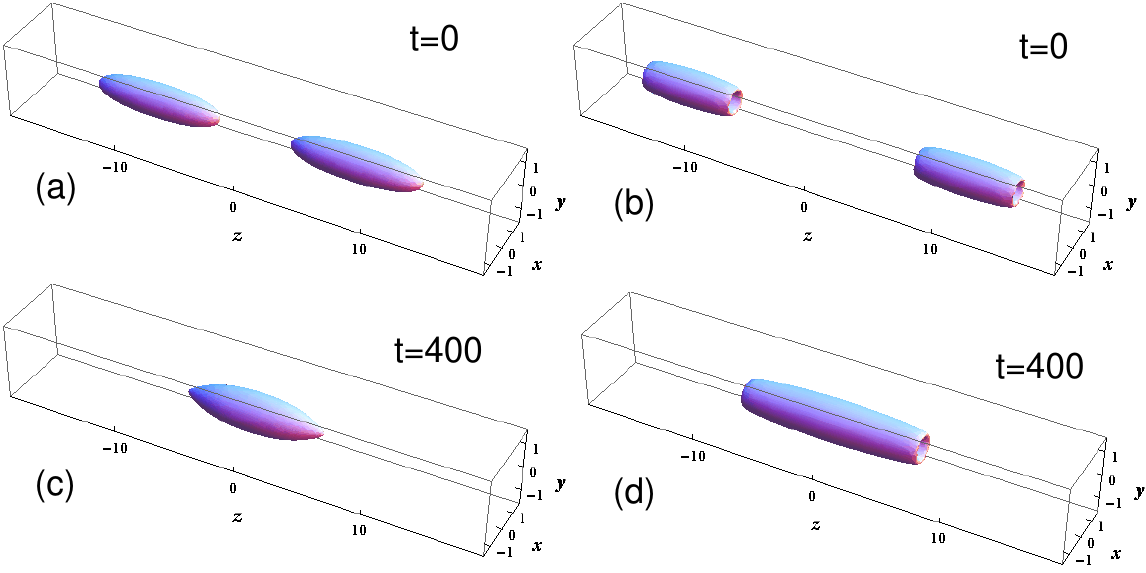}
\end{center}

\caption{{ The initial 3D contour plots of the two (a) bright and (b) vortex solitons 
of Fig. \ref{fig6} at $t=0$ and of the two soliton molecules formed of the  
two (c) bright and (d) vortex solitons at $t=400$ ms. The vortex cores can clearly be 
seen in (b) and (d).
 }}
\label{fig7}
\end{figure}

The  quasi-elastic 
collision at high velocity is  insensitive to the 
initial phase of the solitons. At low velocity, the interaction 
is sensitive to the initial phase. Two solitons placed side-by-side at 
rest along the $z$ direction of opposite phase repeal, 
whereas of the same phase attract and form a stable bound soliton molecule 
\cite{pair}.  

 The dynamics of the formation of a soliton molecule from two 
equal-phase solitons of figure \ref{fig2} is shown in figure \ref{fig6}, 
where we exhibit the contour plot of 1D density $d(z,t)\equiv 2\pi\int \rho 
d \rho |\phi({\bf r},t)|^2$ versus time $t$ and $z$. The initial 
positions of the bright (vortex) solitons are $\pm 17.6$ $\mu$m ($\pm 
22.4$ $\mu$m).  Due to dipolar attraction the solitons come close, 
coalesce and oscillate forming a stable bound soliton molecule \cite{pair}.
{ In the case of the two bright solitons in Fig. \ref{fig6} (a), after formation,
the soliton molecule breaks into two solitons, then forms a molecule, thus oscillating 
between a molecule and two-soliton phases.   
In case of the collision of two vortex solitons, the final  soliton molecule 
is a vortex molecule with zero density along the axial direction. This is illustrated in 
Fig. \ref{fig7} where we show the 3D contour plots of the two interacting bright and vortex solitons
of Fig. \ref{fig6} at $t=0$ and 400 ms.  
}
 A 
pair of bright solitons, bound together by a dark soliton, has been observed 
in dispersion-managed fiber optics \cite{mole}. Now it seems possible to 
observe the present  soliton molecules in DBEC. { It should be possible to consider 
an antivortex \cite{anti} in a DBEC using the present model 
based on a 3D GP equation. A problem of future interest is to consider the collision
between a vortex and antivortex at low velocities. One
may expect that a vortex-antivortex pair will annihilate the angular momentum quantum number,
thus forming after collision a soliton without angular momentum,
instead of forming vortex  molecule of the type shown in Fig.
\ref{fig7} (d) with a hollow along the axial direction.
}
{ Video clips  of the molecule formation dynamics  of figure \ref{fig6}
are also prepared for bright and vortex
solitons  and contained in supplementary clips S3 {\color{red}(also at http://www.youtube.com/watch?v=tnnF0nbi1gs)} 
and S4 {\color{red}(also at http://www.youtube.com/watch?v=01AWIBzbrDs)}, respectively.  }


To conclude, we suggested the possibility of   bright 
and vortex solitons in DBEC for {\it repulsive} short-range interaction 
for $a<a_{dd}$
and studied the statics (rms sizes and chemical potential) and dynamics 
(collision and breathing oscillation) of such solitons. Numerical 
results of stability phase plot and breathing oscillation of the bright 
solitons are  in good agreement with variational 
calculations. The collision between two solitons is 
quasi-elastic at high velocities ($\sim 1$ cm/s)
and independent of initial phase. At very low ($\sim 0$) initial velocity, 
two equal-phase solitons placed side-by-side at rest 
form a bound soliton molecule \cite{pair} due to the dipolar attraction.  
With present technology, these  solitons could be created in 
laboratory and the present theoretical results verified.

\ack
We thank
FAPESP (Brazil),   CNPq (Brazil),    DST (India),   and CSIR  (India)
for partial support.


\end{document}